\documentclass[aps,onecolumn,superscriptaddress,floatfix,nofootinbib,showpacs,altaffilletter]{revtex4-2}

\usepackage{epsfig,graphicx,amssymb,amsmath,wasysym,graphicx}

\usepackage{amsmath}
\usepackage{amsfonts}
\usepackage{amssymb}
\usepackage{graphicx}
\usepackage{bm}
\usepackage{subfigure}
\usepackage{url}
\usepackage[hyperindex]{hyperref}
\usepackage{color}
\usepackage[ddmmyy,24hr]{datetime}
\usepackage{bigdelim}
\usepackage{booktabs}
\usepackage{dcolumn}
\usepackage{multirow}
\usepackage{subfigure}
\usepackage{cancel}
\usepackage{stackrel}
\usepackage{paralist}
\usepackage{xspace}
\usepackage{slashed}
\usepackage{cancel}
\usepackage{todonotes}
\usepackage{enumerate}
\usepackage{float}
\usepackage{fullpage}
\usepackage{ulem}
\usepackage{physics}
\usepackage{orcidlink}
	
\usepackage{lineno}

\newcommand{\nua}[1]{\ensuremath{\rlap{\kern-2.5pt\ensuremath{\overset{\scriptscriptstyle(-)}{\phantom{\nu}}}}{\ensuremath{{\nu}_{#1}}}}}
\newcommand{\vet}[1]{\ensuremath{\hskip-1pt\vec{\hskip1pt#1}}}
\usepackage{enumitem}

\newcommand{\cenns}{CE$\nu$NS\xspace}

\newcommand{\be}{\begin{equation}}
\newcommand{\ee}{\end{equation}}
\newcommand{\ba}{\begin{array}}
\newcommand{\ea}{\end{array}}

\begin{document}

\title{Toward precision physics tests with future COHERENT detectors}

\author{M. Atzori Corona \orcidlink{0000-0001-5092-3602}}
\email{mcorona@roma2.infn.it}
\affiliation{Istituto Nazionale di Fisica Nucleare (INFN), Sezione di Roma Tor Vergata, Via della Ricerca Scientifica, I-00133 Rome, Italy}

\author{M. Cadeddu \orcidlink{0000-0002-3974-1995}}
\email{matteo.cadeddu@ca.infn.it}
\affiliation{Istituto Nazionale di Fisica Nucleare (INFN), Sezione di Cagliari,
	Complesso Universitario di Monserrato - S.P. per Sestu Km 0.700,
	09042 Monserrato (Cagliari), Italy}

\author{N. Cargioli \orcidlink{0000-0002-6515-5850}}
\email{nicola.cargioli@ca.infn.it}
\affiliation{Istituto Nazionale di Fisica Nucleare (INFN), Sezione di Cagliari,
	Complesso Universitario di Monserrato - S.P. per Sestu Km 0.700,
	09042 Monserrato (Cagliari), Italy}

\author{F. Dordei \orcidlink{0000-0002-2571-5067}}
\email{francesca.dordei@cern.ch}
\affiliation{Istituto Nazionale di Fisica Nucleare (INFN), Sezione di Cagliari,
	Complesso Universitario di Monserrato - S.P. per Sestu Km 0.700,
	09042 Monserrato (Cagliari), Italy}

\author{C. Giunti \orcidlink{0000-0003-2281-4788}}
\email{carlo.giunti@to.infn.it}
\affiliation{Istituto Nazionale di Fisica Nucleare (INFN), Sezione di Torino, Via P. Giuria 1, I--10125 Torino, Italy}

\author{R. Pavarani \orcidlink{0009-0004-9534-542X}}
\email{riccardo.pavarani@ca.infn.it}
\affiliation{Istituto Nazionale di Fisica Nucleare (INFN), Sezione di Cagliari,
	Complesso Universitario di Monserrato - S.P. per Sestu Km 0.700,
	09042 Monserrato (Cagliari), Italy}
\affiliation{Università degli Studi di Padova, Dipartimento di Fisica e Astronomia “Galileo Galilei”, Via Francesco Marzolo 8, 35131 Padova, Italy}

\begin{abstract}

We present a comprehensive sensitivity study of future CE$\nu$NS detectors, focusing on a cryogenic cesium iodide detector and a tonne-scale liquid argon one, currently being developed by the COHERENT Collaboration.
These setups will enable precision measurements of the weak mixing angle at low energies and allow accurate extraction of the neutron nuclear distribution radius. We also demonstrate that next-generation detectors will place constraints on the neutrino charge radius comparable to or better than current global fits. In addition, we explore the sensitivity to non standard neutrino electromagnetic properties, such as magnetic moments and millicharges, as well as new mediators. 
These findings reinforce the role of CE$\nu$NS experiments in the upcoming precision era, with future detectors playing a key role in advancing our understanding of neutrino interactions and electroweak physics at low energies.
\end{abstract}

\maketitle

\section{Introduction}
\label{sec:intro}
Predicted over four decades ago~\cite{Freedman:1973yd}, coherent elastic neutrino-nucleus scattering (CE$\nu$NS) is a standard model (SM) neutral-current weak process in which a low-energy neutrino interacts with a nucleus as a whole, via the exchange of a $Z^0$ boson.
Despite its relatively large cross section compared to other low-energy neutrino interactions, the experimental observation of CE$\nu$NS remained elusive for decades due to the extremely low nuclear recoil energies involved. Its first detection by the COHERENT Collaboration~\cite{COHERENT:2017ipa}, using a cesium-iodine detector and neutrinos produced at the Spallation Neutron Source (SNS), marked a milestone in neutrino physics, opening a new avenue for tests of the Standard Model and beyond (BSM)~\cite{DeRomeri:2024hvc,DeRomeri:2024iaw,Pandey:2023arh,Coloma:2023ixt,AristizabalSierra:2024nwf, Cadeddu:2017etk, Cadeddu:2018dux, Cadeddu:2019eta, Cadeddu:2020lky, Cadeddu:2018izq, Cadeddu:2020nbr, Cadeddu:2021ijh, AtzoriCorona:2022moj,AtzoriCorona:2022qrf,AtzoriCorona:2023ktl,AtzoriCorona:2024rtv,Coloma:2017ncl,Liao:2017uzy,Lindner:2016wff,Giunti:2019xpr,Denton:2018xmq,AristizabalSierra:2018eqm,Miranda:2020tif,Banerjee:2021laz,Papoulias:2019lfi,Denton:2022nol,Papoulias:2017qdn,Dutta:2019nbn,Abdullah:2018ykz,Ge:2017mcq,Miranda:2021kre,Flores:2020lji, Farzan:2018gtr, Brdar:2018qqj,AtzoriCorona:2025ygn,Kouzakov:2014lka}. 
Shortly after the observation, COHERENT reported the first observation of CE$\nu$NS on argon nuclei with the CENNS-10 detector~\cite{COHERENT:2020ybo,COHERENT:2020iec}, followed by an updated measurement with significantly improved statistics using the CsI detector~\cite{COHERENT:2021xmm}, which allowed for a more precise determination of the CE$\nu$NS cross section.

More recently, the Collaboration reported the detection of CE$\nu$NS on germanium using the Ge-Mini detector~\cite{COHERENT:2025vuz}, which intriguingly shows a $\sim 2\sigma$ deficit in the measured cross section with respect to its SM prediction, which still lacks a clear physical explanation~\cite{AtzoriCorona:2025xgj}.
Beyond accelerator-based sources, CE$\nu$NS has also been searched for at nuclear power plant sites, where the lower neutrino energies provide a unique environment for testing nuclei at the full coherence regime. In particular, the CONUS+ experiment observed CE$\nu$NS using high-purity germanium detectors in close proximity to a commercial reactor~\cite{Ackermann:2025obx}. Alongside, the first hints of CE$\nu$NS from solar neutrinos have emerged in dark matter experiments such as PandaX~\cite{PandaX:2024muv} and XENONnT~\cite{XENON:2024ijk}, which are beginning to probe the so-called neutrino fog~\cite{OHare:2021utq} as a result of their ultra-low backgrounds and large exposures.
These developments and observations underscore the rapid growth of CE$\nu$NS as a flourishing field, with a diverse and expanding experimental landscape. Multiple detectors are currently operational~\cite{nGeN:2025hsd,CONNIE:2019swq,Ricochet:2022pzj,RED-100:2024izi,TEXONO:2024vfk}, and several others are under development or in the planning stages~\cite{NUCLEUS:2019igx,Cruciani:2022mbb,CHANDLER:2022gvg,MINER:2016igy,NEON:2022hbk,Su:2023klh,Abele:2022iml,Simon:2024cat,Pattavina:2020cqc,Baxter:2019mcx}, aiming to further improve sensitivity, explore new targets and to deepen our understanding of neutrino interactions at low energies.
In this landscape, recent advances in detector technology have significantly enhanced the experimental reach of CE$\nu$NS, leading to a renewed experimental program for the future years. In particular, the COHERENT Collaboration has proposed the installation of a cryogenic cesium-iodine detector (COH-CryoCsI)~\cite{COHERENT:2023sol,COHERENT:2022nrm}, which aims to combine a low energy threshold and increased light yield to significantly improve upon current measurements. Additionally, a future tonne-scale liquid Ar detector is planned, with a significantly increased active mass of the detector, estimated at approximately $750\;\rm kg$ of atmospheric argon (AAr)~\cite{COHERENT:2022nrm}. 
The most abundant component in AAr is the stable isotope $^{40}\rm Ar$, produced via electron capture from $^{40}\rm K$. Since its production rate is proportional to the abundance of $^{40}\rm K$, most of the $^{40}\rm Ar$ originates underground and gradually diffuses into the atmosphere.  Atmospheric argon also contains three long-lived radioactive isotopes: $^{37}\rm Ar$, $^{39}\rm Ar$ and $^{42}\rm Ar$, originating by the interaction of cosmic rays with the atmosphere. Among them, $^{39}\rm Ar$ is a pure $\beta$-emitter and constitutes a significant source of low energy background for argon-based detectors, limiting the sensitivity to rare events searches.
This unstable isotope has an activity of $(1.01\pm0.08)\;\rm Bq\;kg^{-1}$~\cite{WARP:2006nsa}, an endpoint of 565 keV, and a half-life of 269 years. 
To reduce such a background, the DarkSide dark matter Collaboration demonstrated that the use of argon from underground reservoirs (UAr) can improve significantly the experimental reach, 
given that its $^{39}\rm Ar$ content is about 1400 times lower than atmospheric levels~\cite{DarkSide:2015cqb}, corresponding to a rate of $7.3\times10^{-4}~\rm Bq\;kg^{-1}$.
Moreover, the DarkSide Collaboration is building the ARIA plant in Sardinia, which constitutes a tall cryogenic distillation column which will permit to further purify the UAr, both from the $\mathrm{^{39}Ar}$ isotope and from other chemical contaminants~\cite{DarkSide-20k:2023grj,DarkSide-20k:2021nia}.
In this work, we will investigate the potentialities of a detector exploiting UAr for CE$\nu$NS searches, which would result in a completely subdominant $^{39}\rm Ar$ background~\cite{ArMagnificent2025}.
Additional improvements are expected given that the systematic uncertainty on the neutrino flux, dominating current measurements, is expected to be strongly reduced by the implementation of a dedicated $\mathrm{D_2O}$ detector~\cite{COHERENT:2021xhx} approaching $4.7(2)\%$ statistical uncertainty after 2(5) SNS-years of operation.\\
In this work, we provide a comprehensive sensitivity study for these detectors to key electroweak and neutrino parameters, both within the SM and in BSM frameworks, assessing the impact of statistical and systematic uncertainties on the achievable constraints\footnote{A compilation of results from CE$\nu$NS and $\nu$ES probes can be found in the LE$\nu$S-fit web page at \url{https://levs-fit.ca.infn.it}.}. Related studies on the sensitivity of cryogenic CsI and argon detectors at COHERENT to non-standard interactions and leptoquark scenarios can be found in Refs.~\cite{Chatterjee:2024vkd,DeRomeri:2023cjt}.

\section{Theoretical Framework and Sensitivities strategy}
\subsection{CE$\nu$NS cross section}
\label{sec:Theo}
The \cenns cross section as a function of the nuclear recoil energy \(T_\mathrm{nr}\) for a neutrino \(\nu_\ell\) (\(\ell = e, \mu, \tau\)) scattering off a nucleus \(\mathcal{N}\), is
\begin{equation}
    \dfrac{d\sigma_{\nu_{\ell}\text{-}\mathcal{N}}}{d T_\mathrm{nr}} = 
    \dfrac{G_{\text{F}}^2 M}{\pi} 
    \left( 1 - \dfrac{M T_\mathrm{nr}}{2 E^2} \right)
    \left( Q^{V}_{\ell, \mathrm{SM}} \right)^2,
    \label{eq:cexsec}
\end{equation}
where \(G_{\text{F}}\) is the Fermi constant, \(E\) is the neutrino energy, \(M\) is the nuclear mass, and $Q^{V}_{\ell, \mathrm{SM}}$ is the weak nuclear charge, which represents the weak coupling of the neutrino with the nucleus and is given by\footnote{We verified that the inclusion of the axial contribution yields a sub-percent effect, which can therefore be safely neglected~\cite{AbdelKhaleq:2024hir}.}
\begin{equation}
    Q^{V}_{\ell, \mathrm{SM}} = \left[ g_{V}^{p}(\nu_\ell) Z F_Z(|\vec{q}|^2) + g_{V}^{n} N F_N(|\vec{q}|^2) \right],
    \label{eq:weakcharge}
\end{equation}
with $Z (N)$ being the number of protons (neutrons) and $F_{Z(N)}\left(|\vec{q}|^2\right)$ the proton (neutron) nuclear form factor which describes the loss of coherence as a function of the momentum transfer $\left( |\vec{q}| \right)$~\cite{AtzoriCorona:2023ktl}.
We employed the analytical Helm parameterization~\cite{Helm:1956zz}, which depends on the corresponding nuclear rms radius, to describe the nuclear form factors. The Helm form factor is practically equivalent to the other two well-known parameterizations, i.e., the symmetrized Fermi~\cite{Piekarewicz:2016vbn} and Klein-Nystrand~\cite{Klein:1999qj} ones. 
While proton rms radii have been precisely measured for a large number of nuclei~\cite{Fricke:1995zz,Angeli:2013epw}, neutron radii are still poorly known. Therefore, we rely on the predictions from nuclear shell models (NSM)~\cite{Hoferichter:2020osn}. Namely, we consider
\begin{align}
R_n(\textrm{Cs}) &= 5.09\;\textrm{fm},\quad R_n(\textrm{I}) = 5.03\;\textrm{fm}, \\
R_n(\textrm{Ar}) &= 3.55\;\textrm{fm}.
\end{align}
Finally, the coefficients $g_{V}^{n}$ and $g_{V}^{p}$ represent the weak neutral-current vector coupling of the neutrino with the neutron and the proton, respectively. Interestingly, $g_{V}^{p}$ depends on the weak mixing angle, a crucial parameter of the electroweak theory. In the SM, the couplings can be evaluated including the contribution of radiative corrections~\cite{AtzoriCorona:2024rtv,AtzoriCorona:2023ktl,Erler:2013xha,PhysRevD.110.030001}, resulting in
\begin{align}
g_{V}^{p}(\nu_{e}) &= 0.0379,\, g_{V}^{p}(\nu_{\mu}) = 0.0297, \\
g_{V}^{p}(\nu_{\tau}) &= 0.0253,\, g_{V}^{n} = -0.5117,
\end{align}
from which one can note that $g_{V}^{p}$ has a small dependence on the flavor of the incoming neutrino due to the contribution of the neutrino charge radius (CR)~\cite{AtzoriCorona:2024rtv}, which is the only nonzero electromagnetic properties of neutrinos, appearing as a radiative correction to $g_{V}^{p}(\nu_{\ell})$~\cite{Giunti:2024gec}. The SM neutrino CR are given by~\cite{Bernabeu:2000hf,Bernabeu:2002nw}
\begin{equation}
\langle{r}_{\nu_{\ell}}^2\rangle_{\text{SM}}
=
-
\frac{G_{\text{F}}}{2\sqrt{2}\pi^2}
\left[
3 - 2 \ln\left(\frac{m_{\ell}^2}{m_W^2}\right)
\right],
\label{eq:cr-sm}
\end{equation}
where $m_{W}$ is the $W$ boson mass and $m_\ell$ the $\ell$-flavor charged lepton mass. It may be convenient to introduce the flavor independent neutrino-proton coupling, $\tilde{g}_{V}^{p}$, by explicitly separating the charge radius contribution, namely
\begin{equation}
    g_V^p=
    \tilde{g}_{V}^{p} - \dfrac{\sqrt{2} \pi \alpha}{3 G_{\text{F}}} \langle r_{\nu_{\ell}}^2 \rangle,\label{eq:gvpNCR}
\end{equation}
with $\alpha$ being the fine-structure constant and $\tilde{g}_{V}^{p}=0.0182$. In this sensitivity study, we also account for the energy-dependence of the radiative correction associated with the neutrino charge radius~\cite{AtzoriCorona:2024rtv}.

\subsection{COHERENT CryoCsI detectors}

The COHERENT experimental program has led the field of \cenns searches in recent years, with further developments expected in the near future. Among the novel developments foreseen, a fundamental ingredient is represented by the ongoing upgrades of the neutrino source. In the near future, the SNS proton beam energy will increase from 1.01 GeV to 1.3 GeV and the beam power will rise to 2 MW, with a power of 1.7 MW already reached during the data taking of the germanium detector~\cite{COHERENT:2025vuz}. As a result, the number of neutrinos per flavor produced for each proton-on-target will increase to a value of 0.123~\cite{PhysRevD.106.032003}, which is the value adopted for our sensitivity studies. Moreover, a second target station is planned for the 2030s, with a final power of 2.8 MW, significantly increasing the neutrino flux for each neutrino flavor compared to the current configuration.
In this work, we focus on studying the sensitivity reach for the 10 kg COH-CryoCsI I and the subsequent 700 kg COH-CryoCsI II cryogenic CsI detectors, which represent two consecutive steps for the CsI experimental program. We assume a 2 MW beam power for COH-CryoCsI I and a 2.8 MW beam power for COH-CryoCsI II in our analysis, and that the detectors are located at around 19 meters from the SNS source.
We assume a light yield of $50\;\rm PE/ keV_{ee}$, where PE stands for photoelectrons, for both COH-CryoCsI I and COH-CryoCsI II, which represents the highest measurement obtained in several tests performed by the COHERENT Collaboration~\cite{COHERENT:2023sol}. This value is significantly higher than the $13.35\;\rm PE/ keV_{ee}$ achieved by the current COHERENT CsI detector, primarily due to the transition from standard photomultiplies (PMTs) to silicon photomultipliers (SiPMs)~\cite{10.1117/1.OE.53.8.081909}. The behavior of the energy efficiency near the threshold is not well known, as no direct measurement is available. To account for this uncertainty, we set the threshold using a stepping function at $0.8\;\rm{keV}_{\rm nr}$, i.e. $\sim$6 PE, based on the expectation that the acceptance plateau will be reached in that region. These conservative assumptions mitigate potential uncertainties in the knowledge of the acceptance shape near the threshold. 
We adopt the same arrival time distributions used in the current CsI~\cite{COHERENT:2020ybo} detector. The increase in light yield, beyond lowering the threshold, will also improve both detector's time and energy resolutions~\cite{COHERENT:2023sol}. For the purposes of this study, we neglect the effect of the energy resolution. 

A key ingredient to calculate the expected \cenns rate is the nuclear quenching. Generally, only a fraction of the energy deposited by a recoiling nucleus produces scintillation light and is observed. 
The quenching factor might depend on several parameters, including material composition, doping, and temperature~\cite{COHERENT:2021pcd,Lewis:2021cjv}. While data analysis is underway, preliminary estimates point to a roughly energy independent quenching factor of $15 \pm 1.5\%$ as reported in Ref.~\cite{COHERENT:2023sol}.
The COHERENT Collaboration is currently developing a detailed background model which includes events from intrinsic contaminants, afterglow effects in the detector, and external sources, by means of extensive simulations. 
In Fig.~9 of Ref. \cite{COHERENT:2023sol}, the COHERENT Collaboration presents a first tentative estimation of the background obtained by rescaling and extending the background of the current CsI detector, compared to the expected event rate. This can be considered as a very conservative estimation of the background level for the future cryogenic detector as the actual background may be significantly reduced compared to the first CE$\nu$NS measurements~\cite{COHERENT:2023sol}. 
Thanks to the detector's low energy threshold, increased light yield, and enhanced quenching factor, the average signal-to-background ratio (S/B) is $\sim\;3$ within the selected region of interest, and it increases up to $\sim\;7$ in the bins where most of the signal events are expected (see Fig.~8 of Ref.~\cite{COHERENT:2023sol}).
Moreover, since the dominant steady-state background (SSB) is expected to be nearly flat in both energy and time, whereas the signal exhibits a distinct spectral and temporal shape, the overall sensitivity is only weakly dependent on the actual background rate.  
For completeness, in this sensitivity study we include the SSB background contribution, parameterized as a nearly flat background, both in energy and in time, featuring an overall S/B$\sim$3 in the region of interest.  This corresponds to approximately $12\;\rm events/10\;PE$ as extracted from Fig. 8 of Ref.~\cite{COHERENT:2023sol}.
We verified that our framework successfully reproduces the results reported in Ref.~\cite{COHERENT:2023sol} for the selected benchmark models.
Here, we adopt the Asimov dataset~\cite{Cowan2011} to evaluate the test statistic with the most likely dataset (i.e setting all bin contents to their non-integer SM expectation value), which gives the median of the test statistic. To derive the sensitivity on the parameters of interest, we perform a $\chi^2$ analysis, considering both energy and time distributions, defined by\footnote{Given the high statistics expected, this is equivalent to perform a log-likelihood ratio test~\cite{ParticleDataGroup:2024cfk} to derive the sensitivity.}
\begin{equation}
    \chi^2\!=\!\sum_i^{11}\!\sum_j^{30}\!\Bigg[\frac{\Big(1\!+\eta_{f}\!+ \eta_{QF}\Delta^{QF}_{ij}\Big)N_{ij}^{\mathrm{CE}\nu\mathrm{NS}}\!+\eta_{\rm SSB}N_{ij}^{\rm SSB}-\!N_{ij}^{\rm ref}}{\sigma_{ij}}\Bigg]^2+\left(\frac{\eta_{f}}{\sigma_{f}}\right)^2\
    +\left(\eta_{QF}\right)^2+\left(\frac{1-\eta_{\rm SSB}}{\sigma_{\rm SSB}}\right)^2
\end{equation}
where $\sigma_{ij}$ is the statistical uncertainty on the number of events in the $i$th time bin and $j$th energy bin, $N_{ij}^{\rm ref}$, $N^{\rm SSB}_{ij}$ is the number of SSB events expected in each bin, while $N_{ij}^{\mathrm{CE}\nu\mathrm{NS}}$ is the predicted number of events evaluated in the physic scenario under consideration. We consider energy bins of equal width, set to 10 PE, while the time binning follows the same scheme used for the current CsI detector~\cite{COHERENT:2021xmm}.
The nuisance parameters $\eta_{f}$ and $\eta_{QF}$ have been introduced to account for the systematic uncertainties on the neutrino flux and on the quenching factor, respectively. Here, we assume a reference uncertainty of $\sigma_{f}=0.03$, consistent with the estimate adopted in Ref.~\cite{COHERENT:2023sol}, and expected to be achievable within the COHERENT experimental program.
On the other hand, $\Delta^{QF}_{ij}$ quantifies the relative variation in the CE$\nu$NS rate for a $\pm1\sigma$ difference in the quenching factor with respect to its nominal value in each bin (see Fig. 3 of Ref.~\cite{COHERENT:2023sol}). Finally, $\eta_{\rm SSB}$ is the nuisance parameter accounting for the systematic uncertainty on the SSB background component. Its uncertainty is set to $\sigma_{\rm SSB} = 0.02$, consistent with the value obtained for the CsI detector~\cite{COHERENT:2021xmm}.

\subsection{Argon 750 kg detector}

\begin{figure*}[tb]
    \centering
    \includegraphics[width=0.45\linewidth]{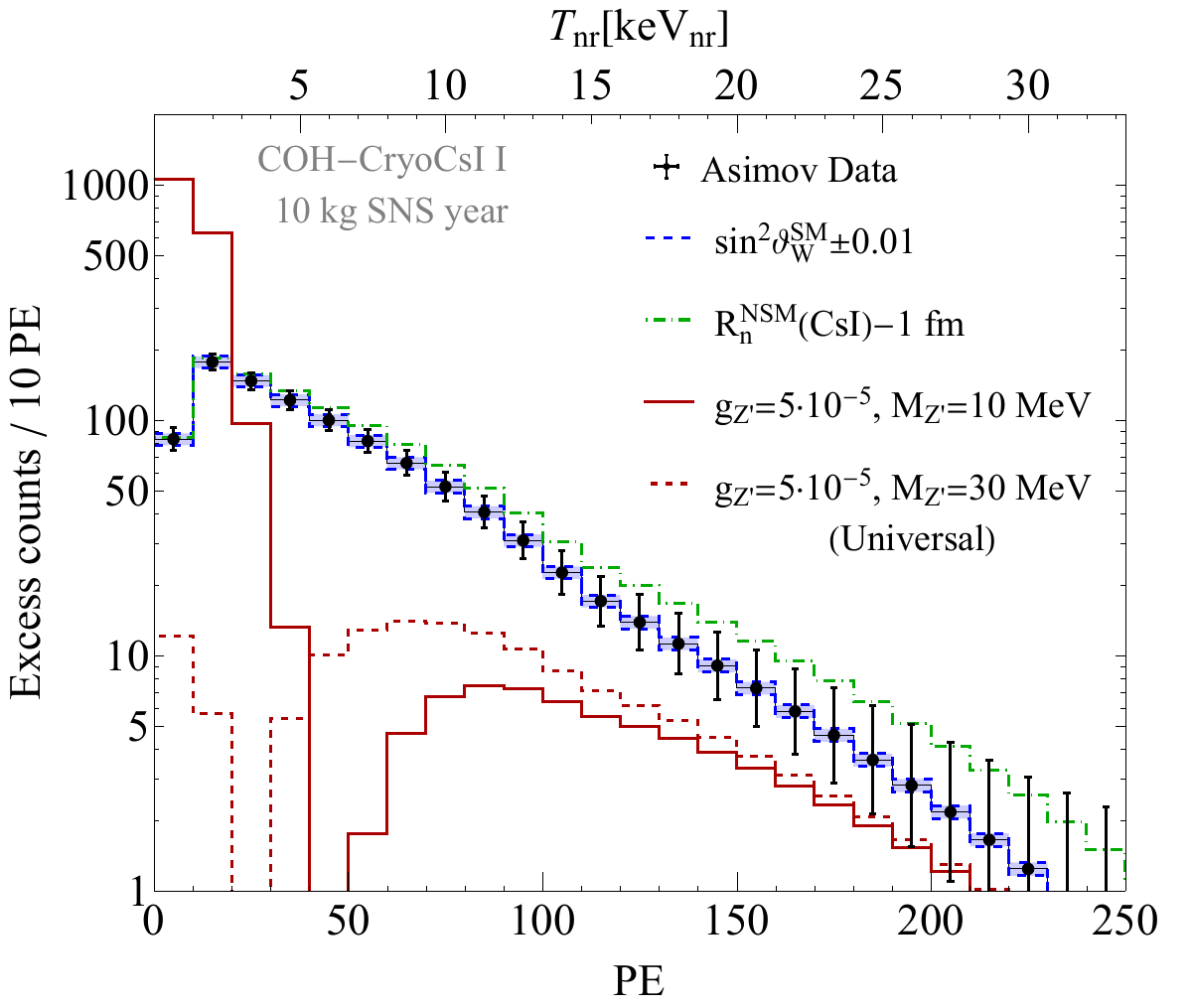}
    \includegraphics[width=0.45\linewidth]{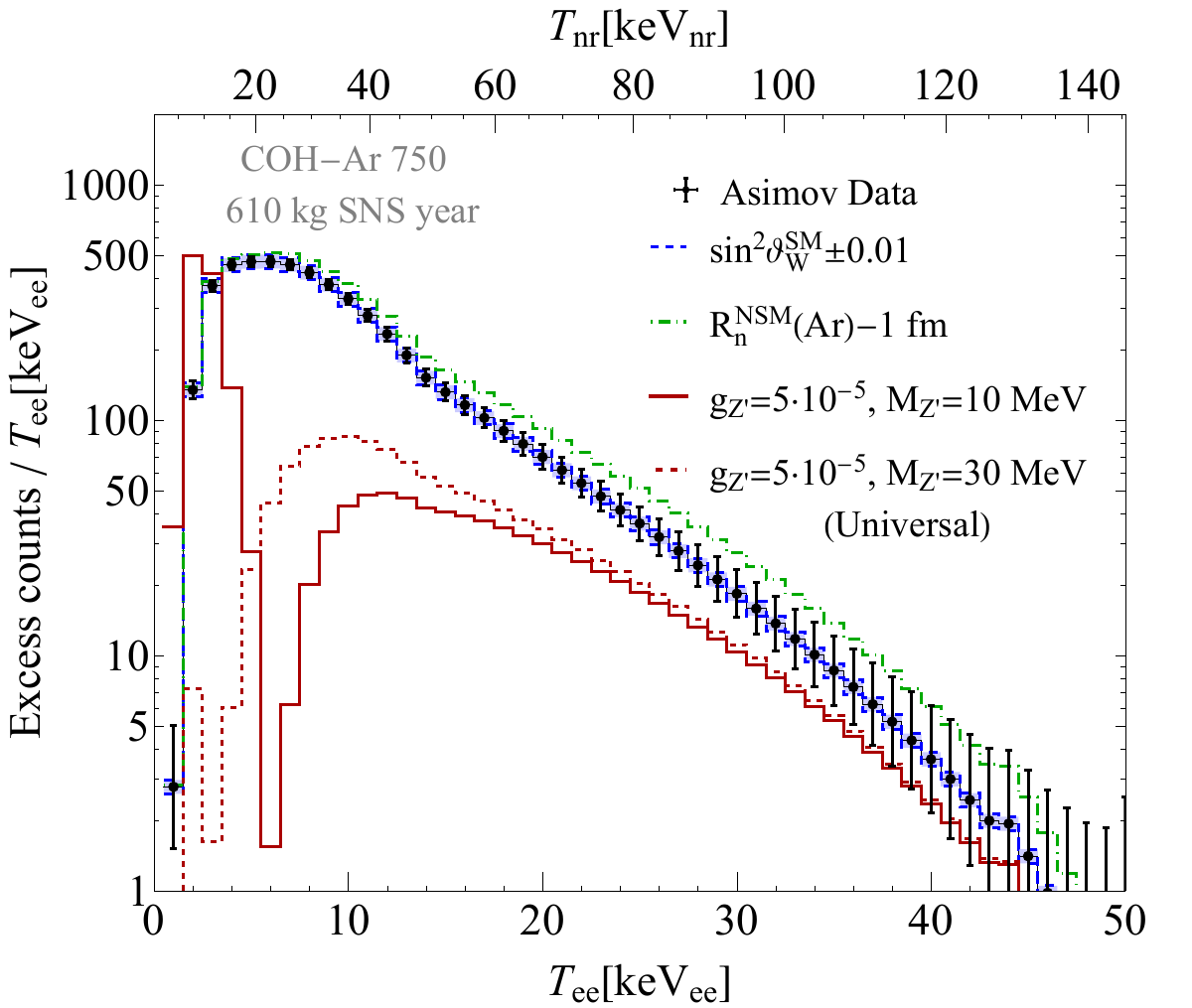}
    \caption{Asimov data for the COH-CryoCsI I (left) and COH-Ar 750 detectors (right) as a function of the recoil energy compared to the expected \cenns rates for different physics scenarios. In blue the event rate for a variation of the weak mixing angle, in green for a different value of the neutron nuclear radius and in red considering the effect of a BSM light mediator universally coupling with the SM fermions, considering the effect of two different mediator masses.}
    \label{fig:DataVsModel}
\end{figure*}

In parallel, the COHERENT Collaboration is also upgrading its liquid argon experimental program, with ongoing efforts toward the development of a tonne-scale liquid argon detector, commonly referred to as the COH-Ar-750 detector. This will employ a cylindrical assembly of PMTs and wavelength-shifting panels reading out a 610 kg active-volume of argon. Current generation cryogen-compatible PMTs have achieved remarkably high single-photon detection efficiency, and we thus assume a $20~\mathrm{keV}_{\rm nr}$ threshold, which is necessary to efficiently discriminate nuclear recoils from electronic recoils~\cite{COHERENT:2022nrm}. Moreover, we consider the same energy acceptance as the CENNS-10 detector from data release~\cite{COHERENT:2020ybo}, and we assume the detector to be located in the same site of the CENNS-10 detector, i.e. 27.5 m. 
Considering time and energy resolution to be well under control, we assume both to be equal to unity.
Similarly to CsI detectors, the energy observed in the CENNS-10 detector, and likewise in the COH-Ar-750, is the electron-equivalent recoil energy $T_\mathrm{ee}$, which is transformed into the nuclear recoil energy through the quenching factor, which is given in Ref.~\cite{COHERENT:2020ybo}.
Concerning the background contributions, the most relevant one is produced by the electron scattering from the decay of $\mathrm{^{39}Ar}$~\cite{ArMagnificent2025}.
The presence of a prominent $\mathrm{^{39}Ar}$ background makes the interpretation of CE$\nu$NS signal challenging, as the determination of a reliable energy spectrum requires knowledge on the so-called $f_\mathrm{prompt}$ function (and its associated systematic uncertainty), which is employed in liquid argon detectors to discriminate efficiently electron recoils from nuclear recoils signals.\footnote{In Ref.~\cite{ATZORICORONA:2025oqh}, a sensitivity study considering AAr was performed by scaling the background observed during the CENNS-10 data taking according to the mass and exposure of the COH-Ar-750 detector. However, given the new design of the detector, such a simple rescaling may introduce biases in the analysis, as the fraction of AAr in the CE$\nu$NS region of interest is expected to be strongly sensitive to the details of the experiment and data analysis.}
To show the potentialities of an alternative approach, we consider a detector filled with UAr, for which $\mathrm{^{39}Ar}$ background rate is obtained by reducing up to a factor $\sim1400$~\cite{DarkSide:2015cqb} the $\mathrm{^{39}Ar}$ content extrapolated from Ref.~\cite{ArMagnificent2025}. This scenario corresponds to a practically background free detector. 
As done for the cryogenic CsI detectors, we perform a sensitivity study adopting Asimov data to evaluate the test statistic with the most-likely dataset and defining the following least-square function
\begin{equation}
\chi^2\!=\!\sum_i^{10}\!\sum_j^{40}\!\left[\!\frac{(1\!+\!\eta_{s})N_{ij}^\mathrm{CE \nu NS}\!-\!N_{ij}^{\rm ref}}{\sigma_{ij}}\!\right]^{\!2}
    + \left(\frac{\eta_{s}}{\sigma_{s}}\right)^2\!.
\end{equation}
In this sensitivity study, we consider energy bins of equal width, set to 1 $\rm{keV}_{ee}$, while the time binning follows the same scheme used for the current CENNS-10 detector~\cite{COHERENT:2020iec,COHERENT:2020ybo}.
The nuisance parameter $\eta_s$ takes into account the total systematic uncertainty on the prediction of \cenns signal and incorporates the systematic uncertainty on the neutrino flux ($\sigma_f=0.03$) and a further contribution to account for all other possible systematic contributions. 
Such systematic effects are difficult to predict with the current information, and thus we consider two scenarios: a conservative one with total systematic uncertainty fixed to $\sigma_s=8\%$, and a more optimistic one with $\sigma_s=5\%$. 

\section{Results}
\label{sec:results}
\begin{figure*}[tb]
    \centering
    \includegraphics[width=0.44\linewidth]{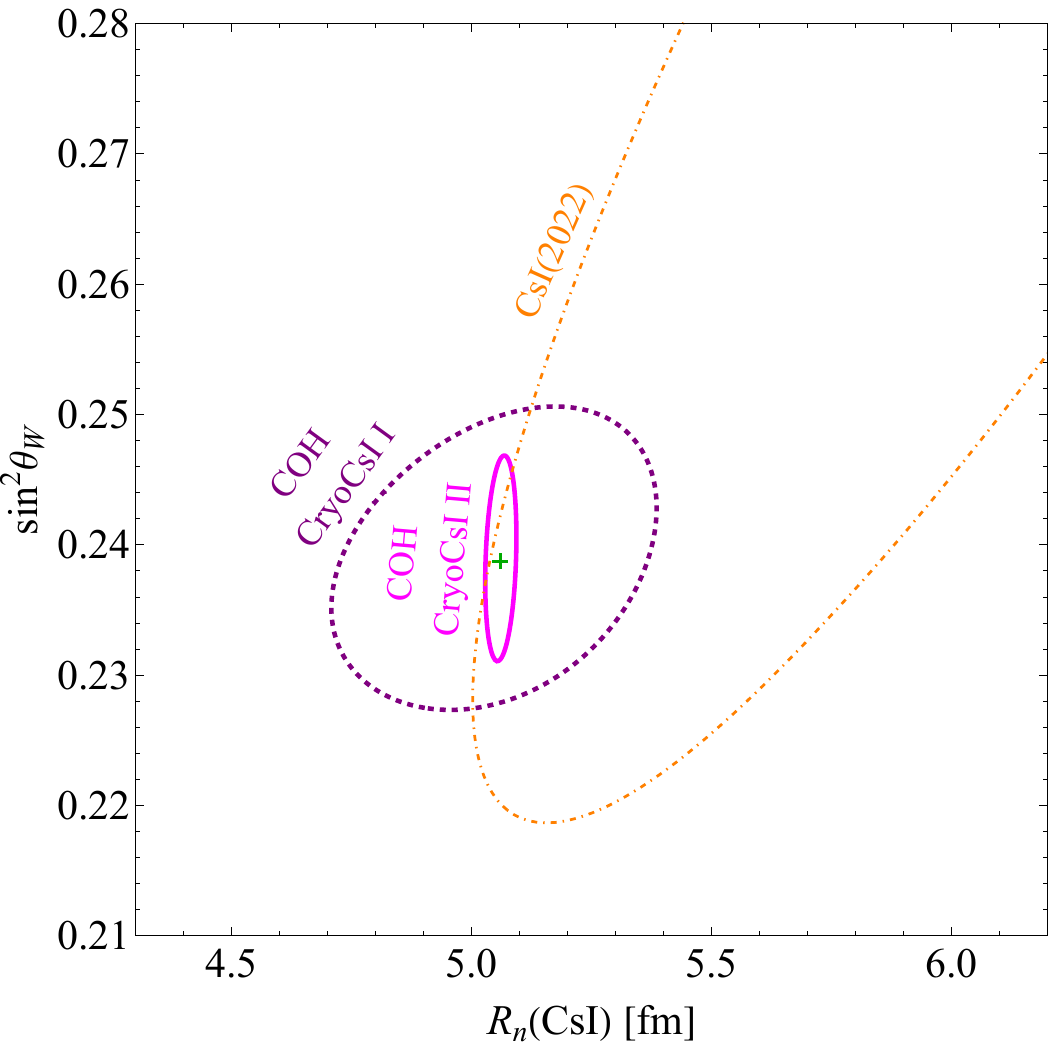}
    \includegraphics[width=0.45\linewidth]{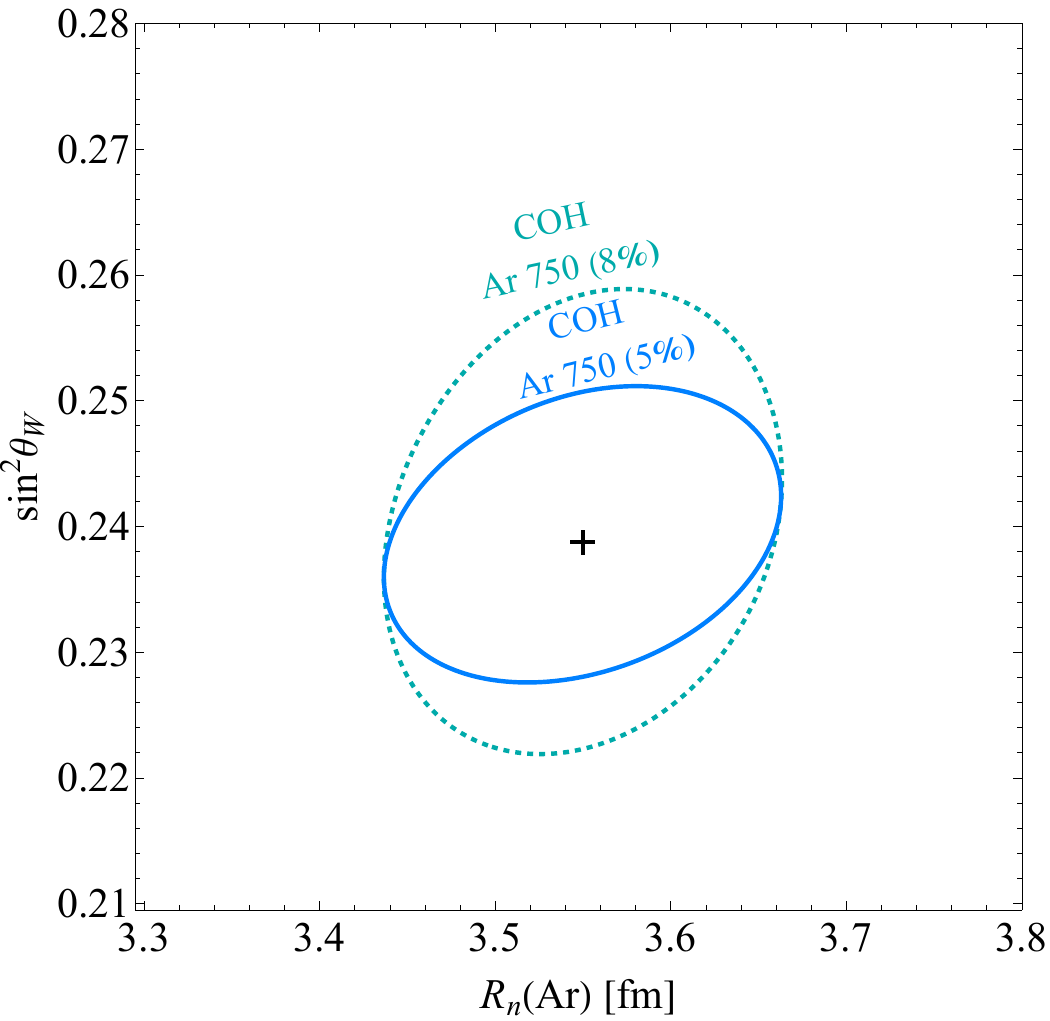}
    \caption{Left: allowed contours at $1\sigma$ CL in the plane of the weak mixing angle and the average neutron radius of cesium and iodine from the current CsI dataset~\cite{AtzoriCorona:2023ktl} compared to the expected sensitivity for the COH-CryoCsI I and \mbox{COH-CryoCsI II} detectors. Right: sensitivity for the future COH-Ar-750 detector assuming two different systematic uncertainties on the CE$\nu$NS signal. The black cross depicts the reference values employed for both parameters, namely $R_n(\textrm{CsI})=5.06 \,\textrm{fm}$, $R_n(\textrm{Ar})=3.55 \,\textrm{fm}$ and $\sin^2\vartheta_W=0.23863$.}
    \label{fig:2Dsin2Rn}
\end{figure*}

In this section, we present the results of our sensitivity analysis for both the COH-CryoCsI I and COH-CryoCsI II detectors and for the COH-Ar 750 detector, considering a variety of different physics models to investigate both SM parameters as well as potential constraints on new physics scenarios. \\
In Fig.~\ref{fig:DataVsModel}, we show the expected data for these detectors along with the expected rates for a different value of the weak mixing angle, for a different value of the nuclear neutron radius and in the presence of a light BSM mediator which couples universally to the SM fermions. It is clear that depending on the physical parameter under investigation, some scenarios produce a shape distortion, others mainly affect the overall normalization, while some may lead to significant enhancements or reductions due to interference effects.
Thus, the extraction of some parameters will be more affected by the improvement of the systematic uncertainties on the CE$\nu$NS signal, though, for others, such improvements will be less relevant. The sensitivities presented here have been obtained considering three SNS years, with one SNS year corresponding to about $\sim5000$ hours of constant operation. According to our calculation, we foresee that the COH-CryoCsI I (COH-Ar 750) detector will observe $1005 \; (5440)\;[\textrm{events}\cdot\textrm{SNS year}]^{-1}$.  
\subsection{Weak mixing angle and neutron distribution radius}
\begin{figure}[!h]
        \centering
        \includegraphics[width=0.45\textwidth]{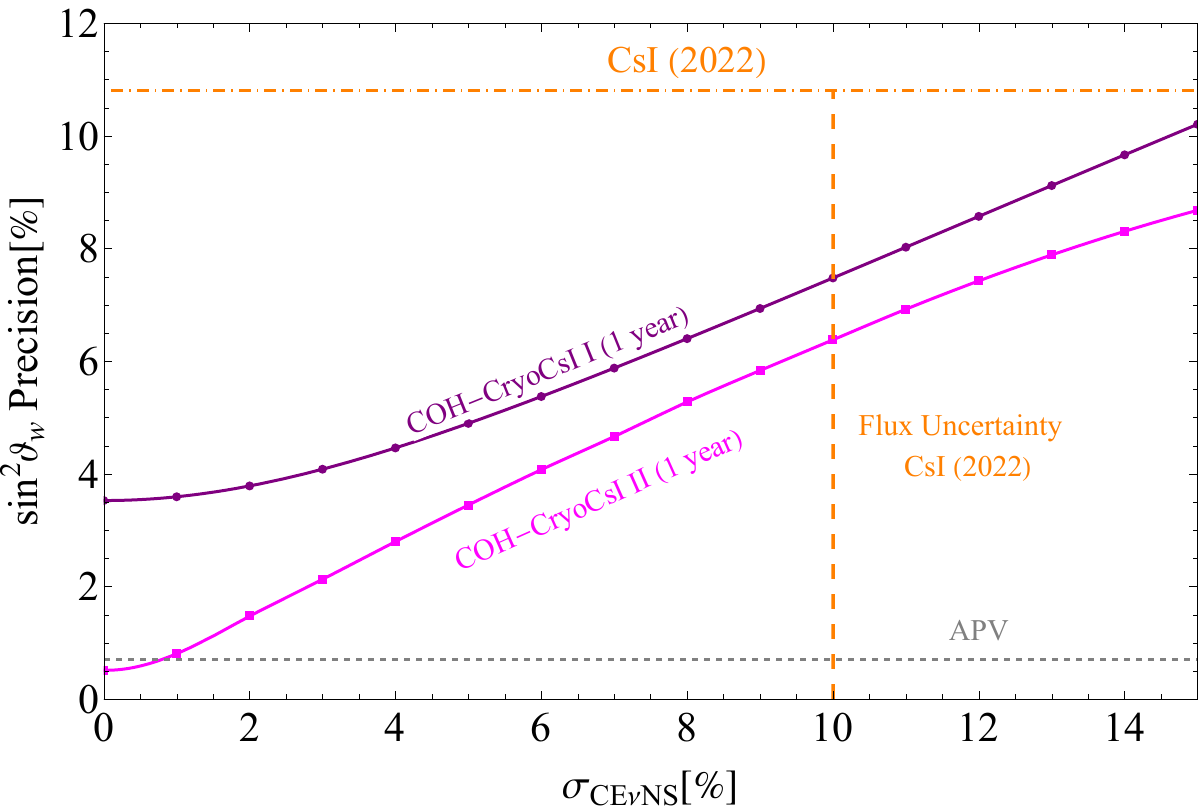}
        \caption{Relative precision on the determination of $\sin^2\theta_W$ as a function of the \cenns systematic uncertainty ($\sigma_{\mathrm{CE\nu NS}}$), for the COH-CryoCsI I and II detectors achievable within 1 SNS year of data taking. The results are compared with the relative uncertainties obtained from the current CsI detector~\cite{AtzoriCorona:2023ktl} and atomic parity violation on cesium~\cite{ParticleDataGroup:2024cfk}.}
        \label{fig:UncSin2cryo}
    \end{figure}

The first parameter that we have investigated is the weak mixing angle, $\theta_{W}$, which being a fundamental parameter of the electroweak theory represents a unique way for testing the SM. Its value is predicted to vary with the energy scale, thus, making it fundamental to measure it at different momentum transfers. Moreover, given that in some BSM scenarios its running can be significantly modified~\cite{PhysRevD.104.L011701}, a deviation of its value from its SM prediction may indicate the presence of new physics.\\
Luckily, \cenns shows the capability of performing measurements of the weak mixing angle for $Q<100\, \mathrm{MeV}$, where measurements are poor, although the precision level reached so far has been limited by the systematic uncertainty on the normalization of the CE$\nu$NS signal. In fact, a shift on $\sin^2\vartheta_W$ produces an almost-constant renormalization of the overall CE$\nu$NS recoil spectrum, as visible in Fig.~\ref{fig:DataVsModel}. 
Furthermore, the \cenns cross section, as clear from Eq.~(\ref{eq:cexsec}), does not depend only on the weak mixing angle value, but also on the neutron rms radius $R_n$ of the target nucleus entering the neutron form factor. The latter is typically poorly known due to the limited availability of probes capable of precisely accessing the neutron density distribution and enabling a model-independent determination. Such a dependence represents a characteristic degeneracy of electroweak probes~\cite{Corona:2021yfd,AtzoriCorona:2023ktl,Cadeddu:2021ijh,AtzoriCorona:2024vhj}. Thus, it is important to extract simultaneously $\sin^{2}{\theta_W}$ and $R_n$ from the data, to avoid misinterpretations.\\
The results of the sensitivity on the simultaneous extraction of the weak mixing angle and the average neutron radius of the cesium and iodine nuclei and the neutron radius of the argon nucleus, respectively, for the COHERENT CryoCsI I, CryoCsI II and Ar 750 detectors are shown at $1\sigma$ confidence level (CL) in Fig.~\ref{fig:2Dsin2Rn}. We compare the sensitivity for the cryogenic CsI detectors with the current precision from COHERENT CsI data~\cite{COHERENT:2021xmm}, from which it is clear that with the future detectors it will be possible to perform such a simultaneous extraction reaching high precision on the neutron radius. Numerically, at $1\sigma$ CL, the results read\footnote{The quoted central values correspond to the input parameters used to generate the Asimov spectra in the sensitivity study.}
\begin{align}
    &\mathrm{Cryo\, I}\!:\!\, \sin^2\!{\vartheta_W}\!=\!0.239^{+0.008}_{-0.007},\,R_n\!=\!5.06^{+0.23}_{-0.22}~\mathrm{fm},\\
    &\mathrm{Cryo\, II}\!:\! \:\sin^2\!{\vartheta_W}\!=\!0.239\!\pm\! 0.005,\,R_n\!=\!5.06\!\pm\! 0.02~\mathrm{fm}\,.
\end{align}
The results of the sensitivity should be compared to the current COHERENT CsI precision of $\sin^2{\vartheta_W}=0.31^{+0.08}_{-0.07}$ and \mbox{$R_n=6.6^{+1.4}_{-1.1}~\mathrm{fm}$} obtained from a simultaneous fit of both parameters~\cite{AtzoriCorona:2023ktl}, and correspond to a relative precision of approximately 25\% for $\sin^2{\vartheta_W}$ and 21\% for $R_n$.
Moving from \mbox{COH-CryoCsI I} to COH-CryoCsI II brings little improvement on the extraction of $\sin^2{\vartheta_W}$ which will be measured to a precision of about 3\%, while it is much more relevant for the neutron radius, whose precision is expected to move from 3\% for COH-CryoCsI I to sub-percent precision for COH-CryoCsI II.
At such a level of precision, it will be possible to disentangle the different contributions of cesium and iodine to the \cenns event rate, which for current data cannot be distinguished. At this level of precision, it will become fundamental to properly consider the nuclear structure contributions separately.\\
The precision on the extraction of $\sin^2{\vartheta_W}$ is mainly determined by the systematic uncertainty rather than the statistics. To show this more quantitatively, in Fig.~\ref{fig:UncSin2cryo}, we show the precision that could be reached by both COH-CryoCsI I and COH-CryoCsI II detectors depending on the total systematic uncertainty. If the latter would be reduced up to roughly 1\%, the \mbox{COH-CryoCsI II} detector could reach the same precision as the measurement of atomic parity violation (APV) on cesium~\cite{doi:10.1126/science.275.5307.1759,PhysRevLett.109.203003}.
    
\begin{figure}[!ht]
        \centering
        \includegraphics[width=0.45\textwidth]{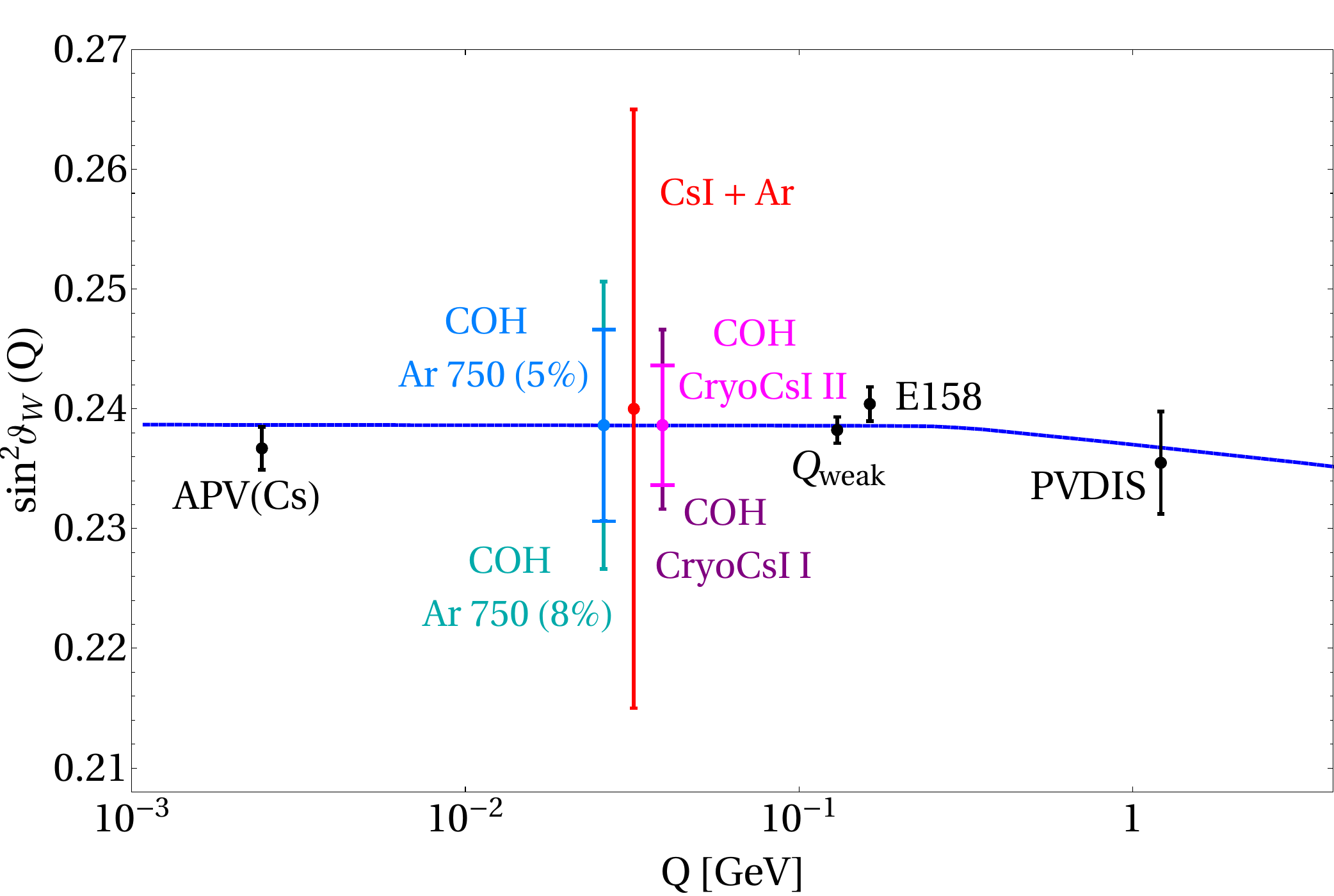}
        \caption{Running of $\sin^{2}{\theta_W}(Q^2)$ with the energy scale Q as predicted by the SM (dotted blue curve), together with experimental determination from atomic parity violation (APV) on cesium~\cite{doi:10.1126/science.275.5307.1759,PhysRevLett.109.203003}, Møller scattering (E158)~\cite{PhysRevLett.95.081601}, deep inelastic scattering of polarized electrons on deuterons (PVDIS)~\cite{Wang2014} the result from the proton's weak charge ($Q_\mathrm{weak}$)~\cite{2018} and the combined analysis of COHERENT CsI and Ar data sets~\cite{DeRomeri:2022twg,AtzoriCorona:2025xgj}.
        In purple (magenta), our result for the COH-CryoCsI I (COH-CryoCsI II) detector is shown, while the darker cyan (azure) data point corresponds to the COH-Ar-750 detector result in the conservative (optimistic) scenario.}
        \label{fig:runningweak}
    \end{figure}

The results obtained for the Ar 750 detector under the two different systematic uncertainty scenarios lead to 
\begin{align}
    &\mathrm{Ar\,750\:(8\%)}\!:\!\, \sin^2\!{\vartheta_W}\!=\!0.239^{+0.013}_{-0.012},\,R_n\!=\!3.55^{+0.07}_{-0.08}~\mathrm{fm},\\
    &\mathrm{Ar\,750\:(5\%)}\!:\! \:\sin^2\!{\vartheta_W}\!=\!0.239^{+0.008}_{-0.007},\,R_n\!=\!3.55^{+0.07}_{-0.08}~\mathrm{fm}\,,
\end{align}
from which it is visible that the reduced systematic uncertainty improves the precision on the weak mixing angle, while leaving the neutron radius practically unaffected. Notably, the results show a significant improvement compared to those from \mbox{CENNS-10}~\cite{ATZORICORONA:2025oqh,Cadeddu:2020lky,COHERENT:2020ybo}, which allowed to obtain only an upper limit on the neutron radius of $4.4\;\mathrm{fm}$ ($1\sigma$), while the precision on the weak mixing angle was around 20\%.
\\
In Fig.~\ref{fig:runningweak}, we compare the precision on the weak mixing angle as found by our sensitivity study with the other available measurements in the low energy regime, along with the SM predicted running of $\sin^2{\vartheta_W}$, calculated in the $\overline{\mathrm{MS}}$ renormalization scheme~\cite{10.1093/ptep/ptac097,PhysRevD.72.073003,Erler_2018}. As it can be observed, 
despite having a worse precision with respect to APV, at least for the assumed level of systematic uncertainties,
it will improve significantly with respect to current \cenns experimental precision~\cite{AtzoriCorona:2023ktl,Alpizar-Venegas:2025wor,DeRomeri:2025csu,DeRomeri:2022twg,DeRomeri:2024iaw}, contributing to fill the gap of weak mixing angle measurements at low energies.

\subsection{Neutrino charge radii}

\begin{figure}[!ht]
        \centering
        \includegraphics[width=0.45\textwidth]{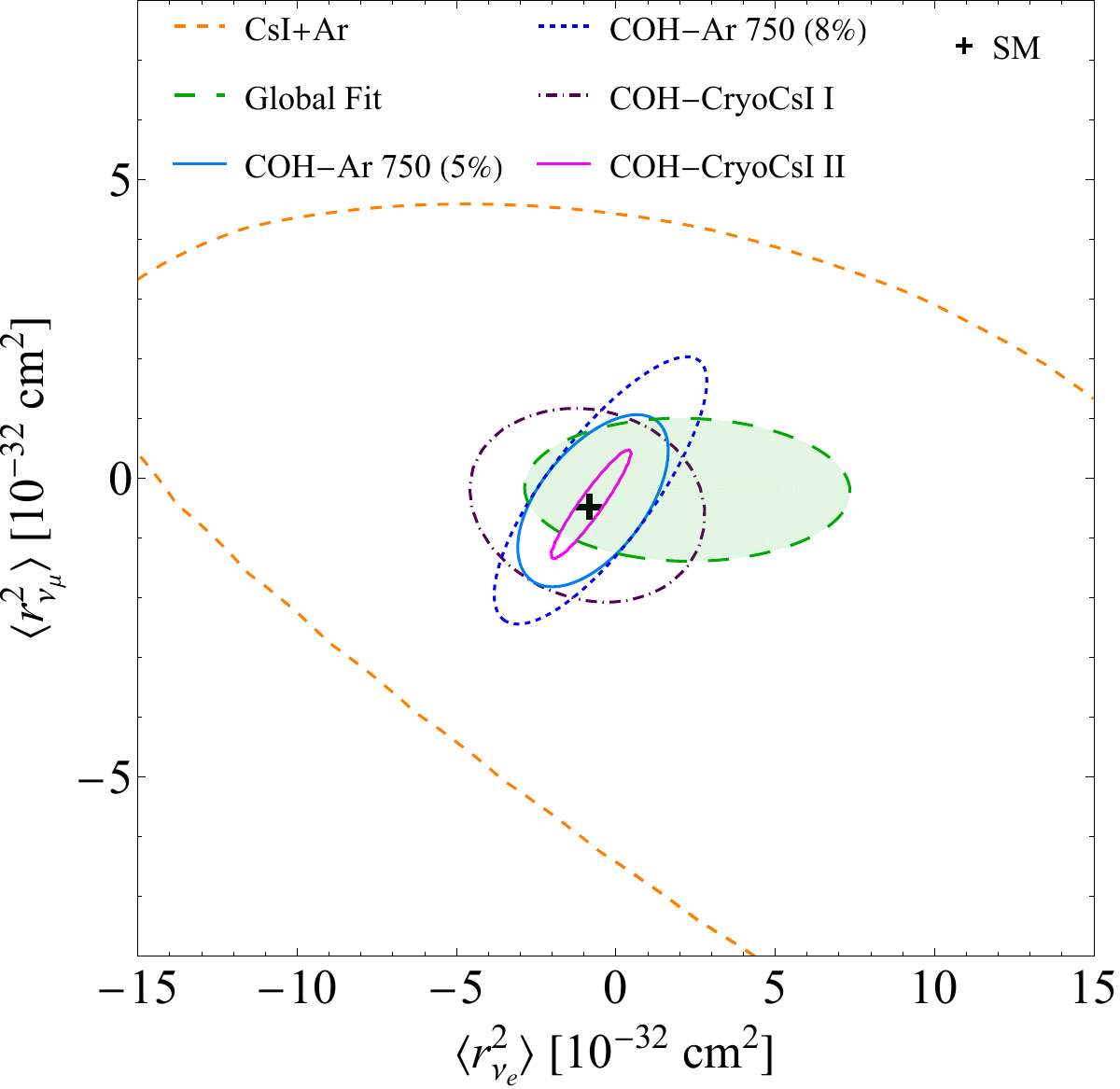}
        \caption{Allowed contours at $90\%$ CL in the plane of the muon and electron neutrino charge radii obtained by our sensitivity study for the COH-CryoCsI I, COH-CryoCsI II and Ar-750 detectors, compared to the combined result from the current COHERENT CsI and Ar data (orange)~\cite{AtzoriCorona:2024rtv}, and the result from a recent global fit of \cenns and neutrino-electron scattering data (green contour)~\cite{AtzoriCorona:2025xwr}. The black cross depicts the SM prediction.}
        \label{fig:2DreeruuAll}
    \end{figure}

As discussed in the introduction, the \cenns process permits also to investigate the neutrino charge radius, $\langle{r}_{\nu_{\ell}}^2\rangle_{\text{SM}}$, which enters the neutrino-proton coupling as shown in Eq.~(\ref{eq:gvpNCR}). According to the SM, the neutrino charge radius is an intrinsic neutrino property and preserves the neutrino flavor in the interaction. However, in some BSM scenarios, neutrino CR may include small off-diagonal terms~\cite{Kouzakov:2017hbc,AtzoriCorona:2022qrf,Cadeddu:2018dux} (usually called transition CR), $\langle r_{\nu_{\ell\ell'}}^2 \rangle$.  In this work, we will only consider the scenarios including diagonal terms, in order to determine the possible precision that could be reached on such parameters, especially for the muon and electron neutrino's CR, given that COHERENT does not have a tau neutrino flux.
By leaving free-to-vary the muon and electron neutrino CR inside the cross section, we are able to obtain the constraints shown in Fig.~\ref{fig:2DreeruuAll} for the COH-CryoCsI I, COH-CryoCsI II and COH-Ar 750 detectors.
We compare our results with the contours obtained with the current CsI and Ar data~\cite{AtzoriCorona:2024rtv}, and with the results of a global fit of all available \cenns and neutrino-electron scattering data~\cite{AtzoriCorona:2025xwr}. The numerical results at 90\% CL are reported in Table~\ref{tab:NCR}, while the constraints are presented in Fig.~\ref{fig:2DreeruuAll}.
\begin{table}[h!]
\centering
\renewcommand{\arraystretch}{1.5}
\resizebox{0.5\columnwidth}{!}
{
\begin{tabular}{c|c||c}

 & $\langle r_{\nu_e}^2 \rangle\;[10^{-32}\mathrm{cm}^2]$ & $\langle r_{\nu_\mu}^2 \rangle\;[10^{-32}\mathrm{cm}^2]$\\
\hline
 COH-CryoCsI I & [-3.7,2.0] & [-1.7,0.8] \\
\hline
 COH-CryoCsI II &[-2.0,0.2] & [-1.2,0.3]\\
 \hline
 COH-Ar 750 ($8\%$) &[-3.2,1.9] & [-2.0,1.4]\\
 \hline
 COH-Ar 750 ($5\%$) &[-2.6,1.0] & [-1.5,0.7]\\
\end{tabular}}
\caption{Constraints at $90\%$ CL on the electron and muon neutrino charge radii obtained by our sensitivity study for the COH-CryoCsI I, COH-CryoCsI II and Ar-750 detectors.}\label{tab:NCR}
\end{table}
Intriguingly, the next generation of COHERENT detectors will be able to reach a precision level similar or even better than that of the global fit~\cite{AtzoriCorona:2025xwr}, especially on the electron neutrino CR, showing the capability of probing directly the SM prediction without the need of the inclusion of other experimental data. It is also relevant to discuss that current COHERENT data are not able to select the parameter space around the SM prediction alone, but show four allowed regions at $90\%$ CL~\cite{AtzoriCorona:2024rtv} which correspond to a degeneracy in the \cenns cross section. Thanks to the expected precision, the future COHERENT detector will overcome such a problem, selecting only one allowed region around the SM expectation.

\subsection{Neutrino magnetic moments}
Along with the neutrino charge radius, it is possible to investigate other neutrino electromagnetic properties that could arise in many beyond the standard model scenarios. Among them, the neutrino magnetic moment (MM) is the most investigated one, as its existence may arise naturally by considering massive neutrinos~\cite{Giunti:2014ixa,Giunti:2015gga,Giunti:2024gec}.
The MM contribution does not interfere with the SM one, and thus it is accounted for by adding the MM contribution to the SM cross section in Eq.~(\ref{eq:cexsec}), namely
\begin{equation}
\dfrac{d\sigma_{\nu_{\ell}\text{-}\mathcal{N}}^{\text{MM}}}{d T_\mathrm{nr}}
=
\dfrac{ \pi \alpha^2 }{ m_{e}^2 }
\left( \dfrac{1}{T_\mathrm{nr}} - \dfrac{1}{E} \right)
Z^2 F_{Z}^2(|\vec{q}|^2)
\left| \dfrac{\mu_{\nu_{\ell}}}{\mu_{\text{B}}} \right|^2
,
\label{cs-mag}
\end{equation}
where $\mu_{\nu_{\ell}}$ is the effective MM of the flavor neutrino $\nu_{\ell}$~\cite{Giunti:2014ixa}, $\mu_{\text{B}}$ is the Bohr magneton and $m_e$ the electron mass. The MM contribution scales with the inverse of the recoil energy, thus, low threshold experiments have strong sensitivity to this quantity. In this sense, as done in previous works~\cite{AtzoriCorona:2022qrf,Coloma:2022avw,DeRomeri:2022twg}, it is possible to consider the contribution from another neutrino process, namely the elastic scattering of neutrinos off atomic electrons, to set stronger constraints on the neutrino MM. Typically, the neutrino-electron scattering produces a negligible contribution to \cenns experiments, however, thanks to the much lighter electron mass, the MM contribution can enhance significantly the neutrino-electron cross section making its contribution relevant when trying to constrain it. 
\begin{table}[h!]
\centering
\renewcommand{\arraystretch}{1.5}
\resizebox{0.5\columnwidth}{!}
{
\begin{tabular}{c|c|c||c|c}

 & \multicolumn{2}{c||}{$\mu_\nu\;[10^{-10}\mu_B]$} & \multicolumn{2}{c}{$q_\nu\;[10^{-10}e_0]$} \\
\cline{2-5}
 & $|\mu_{\nu_e}|$ & $|\mu_{\nu_\mu}|$ & $q_{\nu_e}$ & $q_{\nu_\mu}$ \\
\hline
 COH-CryoCsI I & $<9.8$ & $<7.4$ & $<1.4$ & $<1.0$ \\
\hline
 COH-CryoCsI II & $<3.2$ & $<2.5$ & $<0.5$ & $<0.3$ \\
  \hline\hline
 COH-Ar 750 ($8\%$) & $<8.4$ & $<6.6$ & $[-42,44]$ & $[-23,24]$\\
 \hline
 COH-Ar 750 ($5\%$) & $<8.2$ & $<6.5$ & $[-42,44]$ & $[-23,24]$ \\

\end{tabular}}
\caption{Constraints at $90\%$ CL on the electron and muon neutrino magnetic moment (left) and millicharge (right) obtained by our sensitivity study for the COH-CryoCsI I, COH-CryoCsI II and COH-Ar-750 detectors. For the CsI detectors, we include the contribution from the neutrino-electron scattering channel.}\label{tab:nuEM}
\end{table}
In our analysis this is taken into account only for the COH-CryoCsI detectors, because the Ar-750 one will discriminate electron recoil signals from nuclear recoil ones, thanks to the pulse shape discrimination.  Similarly to \cenns, the MM neutrino-electron cross section reads
\begin{equation}
\dfrac{d\sigma_{\nu_{\ell}\text{-}\mathcal{A}}^{\text{MM}}}{d T_\mathrm{e}}
=
Z_{\text{eff}}^{\mathcal{A}}(T_{\text{e}}) \dfrac{ \pi \alpha^2 }{ m_{e}^2 }
\left( \dfrac{1}{T_\mathrm{e}} - \dfrac{1}{E} \right)
\left| \dfrac{\mu_{\nu_{\ell}}}{\mu_{\text{B}}} \right|^2,
\label{es-mag}
\end{equation}
with $Z_{\text{eff}}^{\mathcal{A}}$ the effective number of electrons that can be ionized by a $T_e$ energy deposit. For a more detailed discussion of the neutrino-electron cross section please refer to Refs.~\cite{AtzoriCorona:2022qrf,Coloma:2022avw,DeRomeri:2022twg}. According to our sensitivity study, whose numerical results are reported in Tab.~\ref{tab:nuEM}, the future COHERENT cryogenic CsI and LAr detectors will improve significantly with respect to the current CsI and Ar constraints, up to about an order of magnitude~\cite{AtzoriCorona:2022qrf}.
Such limits, however, are not competitive compared to current bounds from reactor~\cite{TEXONO:2006xds,Beda:2012zz,AtzoriCorona:2025ygn} or solar neutrino experiments\footnote{However, as noted in Ref.~\cite{Ternes:2025lqh}, limits from different sources cannot be directly compared, as the effective neutrino magnetic moments considered in this study differs fundamentally when considering solar, accelerator or reactor experiments. The limits reported here are intended to serve as an estimation of the physics reach of the experiment.}~\cite{A:2022acy,AtzoriCorona:2022qrf,Giunti:2023yha,Borexino:2017fbd} (see Fig.~5 of Ref.~\cite{AtzoriCorona:2025ygn}). For reference, present direct constraints on the electron neutrino magnetic moment from reactor experiments reached the level of
$\mu_{\nu_e} \lesssim 2.9\times10^{-11}\,\mu_B,$~\cite{Beda:2012zz}
while solar neutrino measurements improve the limit on the effective solar neutrino magnetic moment down to
$\mu_{\nu_s} \lesssim 6.4\times10^{-12}\,\mu_B$~\cite{XENON:2022ltv}.
A more complete picture of the existing limits can be found in Fig.~5 of Ref.~\cite{AtzoriCorona:2025ygn}. While lowering the detection threshold helps to improve the constraints, the main intrinsic limitation of the SNS compared to reactor or solar experiments lies in the neutrino flux. In fact, reactor and solar sources produce a prominent flux of neutrinos with energies around $\mathcal{O}(1\;\rm{MeV})$, which is absent at the SNS, resulting in a stronger neutrino-electron scattering contribution in the CE$\nu$NS region of interest.
\subsection{Neutrino millicharges}
Furthermore, it is possible to consider the existence of a tiny neutrino electric charge (EC), usually referred to as millicharge. In the latter case, the differential \cenns cross section for a millicharged neutrino can be retrieved by replacing the neutrino proton coupling inside the nuclear weak charge in Eq.~(\ref{eq:weakcharge}) by~\cite{Kouzakov:2017hbc,Giunti:2014ixa}
\begin{equation}
g_{V}^{p}(\nu_\ell)\rightarrow 
g_{V}^{p}(\nu_\ell)-\dfrac{ 2 \sqrt{2} \pi \alpha }{ G_{\text{F}} q^2 }
\, q_{\nu_{\ell}}
,
\label{Qech}
\end{equation}
where $q_{\nu_{\ell}}$ is the neutrino EC. Given that the contribution of the millicharge depends on the inverse of the momentum transfer, also in this scenario, considering the contribution due to neutrino-electron scattering enables to set much tighter constraints.
Similarly to CE$\nu$NS, the contribution of EC is accounted for by replacing the neutrino-electron vector coupling inside the SM cross section by ~\cite{AtzoriCorona:2022jeb}
\begin{equation}
    g_{V}^{\nu_{\ell}} \to 
    g_V^{\nu_\ell}+\dfrac{ 2 \sqrt{2} \pi \alpha }{ G_{\text{F}} q^2 }\, q_{\nu_{\ell}}.
\end{equation}
The numerical constraints are reported in Tab.~\ref{tab:nuEM} at the 90\% CL. While the constraints show an improvement compared to the current CsI and Ar limits~\cite{AtzoriCorona:2022qrf}, they remain weaker than those obtained from other existing bounds, see e.g. Fig.~5 of Ref.~\cite{AtzoriCorona:2025ygn}. The same considerations discussed for the magnetic moment results apply to the electric charge ones. Clearly, since we are not including the neutrino-electron scattering channel in the COH-Ar-750 detector, the resulting constraints are significantly weaker compared to those from the COH-CryoCsI detectors. This conclusion reinforces the importance of considering CE$\nu$NS data from reactors to obtain more competitive and comprehensive constraints.

\subsection{Search For New Mediators}

\begin{figure*}[t!]
    \centering
    \includegraphics[width=0.44\linewidth]{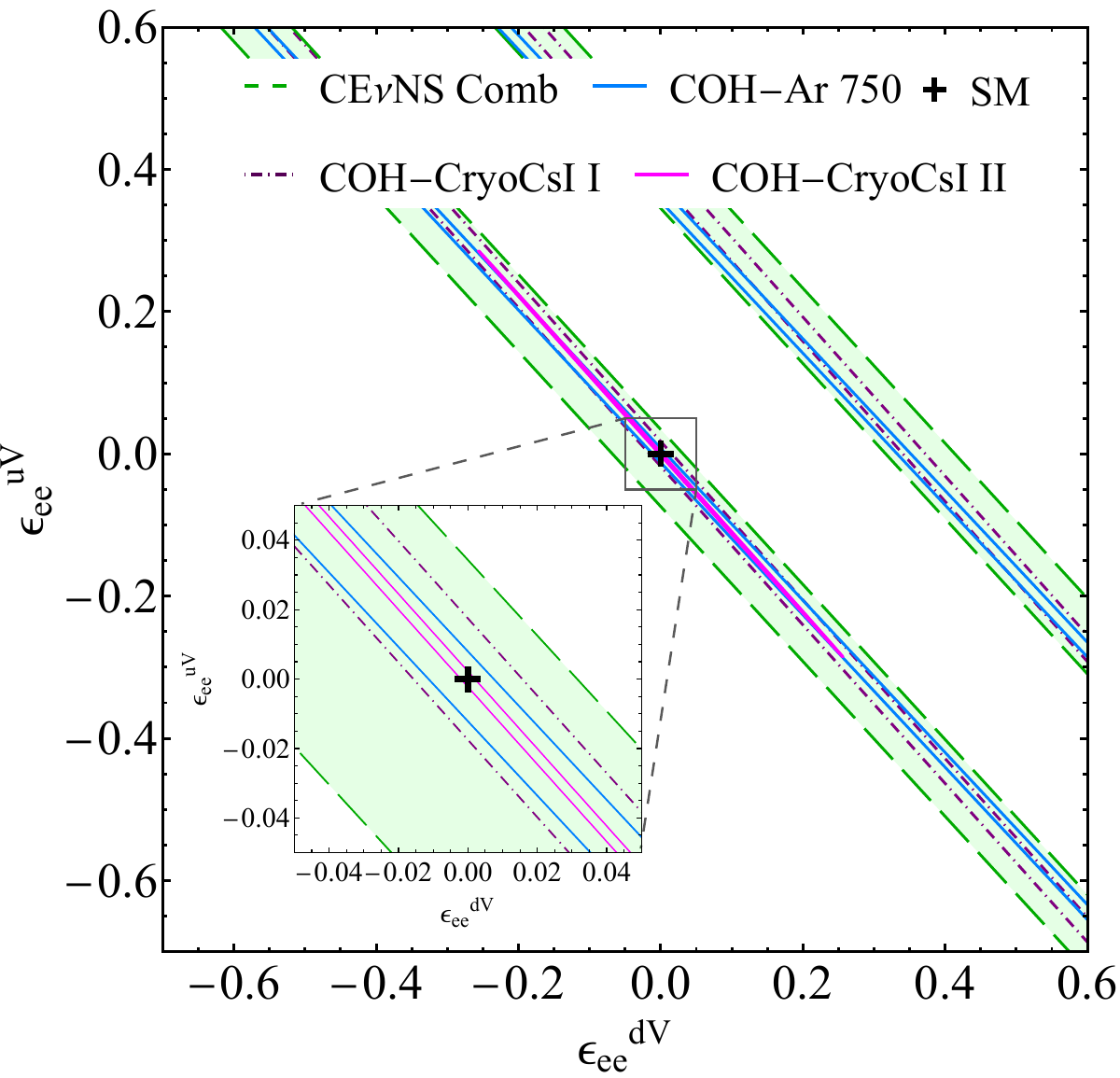}
    \includegraphics[width=0.45\linewidth]{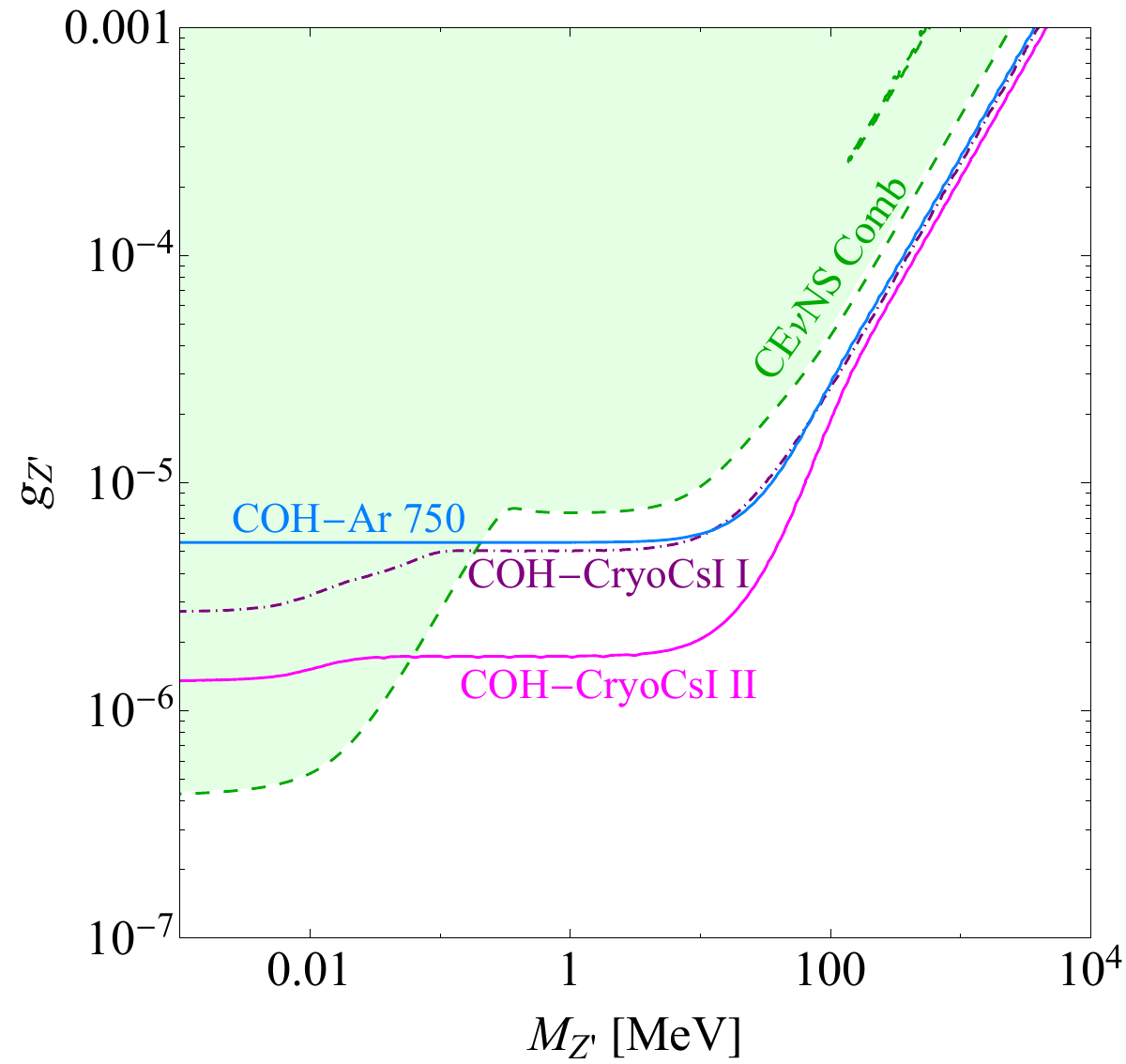}
    \caption{Allowed contours at $90\%$ CL on flavor-preserving NSI (left) and in the plane of the coupling and the mass of a hypothetical universal light mediator obtained by our sensitivity study for the Cryo I, Cryo II and Ar-750 detectors, compared to the current constraint from the combination of other available \cenns experiments presented in Ref.~\cite{AtzoriCorona:2025ygn}. The inset in the left panel shows a zoom in the parameter space to better appreciate the improvement compared to the current limit.}\label{fig:LMlimit}
\end{figure*}  
In this section, we examine the possibility that a new massive vector boson might mediate the neutrino interaction~\cite{Giunti:2019xpr, Coloma:2023ixt}, so-called neutrino nonstandard interactions (NSIs). Under the assumption that the vector boson couples to SM leptons and quarks, the nuclear weak charge in Eq.~(\ref{eq:weakcharge}) is modified by~\cite{Cadeddu:2020nbr,AtzoriCorona:2022moj}
\begin{equation}
Q_{\ell,\mathrm{NSI}}^{V}
=
\left( g_{V}^{p}(\nu_{\ell}) + 2 \varepsilon_{\ell\ell}^{uV} + \varepsilon_{\ell\ell}^{dV} \right)
Z
F_{Z}(|\vet{q}|^2)
+ \left( g_{V}^{n} + \varepsilon_{\ell\ell}^{uV} + 2 \varepsilon_{\ell\ell}^{dV} \right)
N
F_{N}(|\vet{q}|^2),
\label{Qalpha2}
\end{equation}
with $f = u, d$ and $\ell=e,\mu$, and $\varepsilon_{\ell\ell}^{fV}$ representing the size of NSI relative to standard neutral-current weak interaction.
In Fig.~\ref{fig:LMlimit} (left), we show the sensitivity of future COHERENT detectors to the simplified flavor-preserving scenario, which involves only two nonzero NSI parameters, namely $\epsilon^{\mathrm{uV}}_{\mathrm{ee}}$ and $\epsilon^{\mathrm{dV}}_{\mathrm{ee}}$. We compare the expected sensitivity from the COH-CryoCsI I, II and Ar-750 detectors with the constraints from the combined analysis of CE$\nu$NS probes in Ref.~\cite{AtzoriCorona:2025ygn}, finding that the future detectors will significantly improve the constraints. 
Figure~\ref{fig:LMlimit} shows that COHERENT COH-CryoCsI II will be able to reach unprecedented precision in such a scenario, possibly excluding the band degenerate with the SM one. Furthermore, the different inclinations of the Ar and CsI bands can be appreciated, reflecting the use of targets with different $N/Z$ ratios.\\
On the other hand, we can consider the scenario where the new interaction is mediated by a light vector boson, usually referred to as $Z'$. In this case, the NSI parameters take the form of a propagator~\cite{Cadeddu:2020nbr,AtzoriCorona:2022moj},
\begin{equation}
    \epsilon_{\ell \ell}^{fV}=\dfrac{g_{Z'}^2\,Q'_\ell Q'_f}{\sqrt{2}G_F\, (|\vec{q}|^2+M_{Z'}^2)},
\end{equation}
where $m_{Z'}$ represent the mass of the new boson and $g_{Z'}$ is the coupling constant, while $Q'$ are the charges under the new gauge symmetry $U(1)'$.
Since the propagator depends on the momentum transfer, neutrino-electron scattering may also be relevant in the light mediator scenario.
For the latter process, the cross section for the light mediator contribution can be obtained by substituting the neutrino-electron vector coupling with~\cite{DeRomeri:2024dbv,DeRomeri:2025csu}
\begin{equation}
g_V^{\nu_\ell}\rightarrow g_V^{\nu_\ell}+\dfrac{g_{Z'}^2\,Q'_\ell Q'_e}{\sqrt{2}G_F\, (|\vec{q}|^2+M_{Z'}^2)}\,.
\end{equation}
The introduction of the neutrino-electron scattering process, whether possible as in the case of cryogenic CsI detectors, usually produces an enhanced sensitivity to lighter $Z'$ bosons with respect to the \cenns constraints.\\
To provide an exemplification of the possible sensitivity for the future Cryo I, Cryo II and Ar-750 detectors, we consider a simple scenario, for which the $Z'$ boson couples universally to all SM fermions~\cite{Liao:2017uzy,Papoulias:2017qdn,Papoulias:2019txv,Cadeddu:2020nbr,Bertuzzo:2021opb,AtzoriCorona:2022moj}, also known as the universal model. While this model is not gauge invariant and requires right handed neutrinos to be anomaly free, it serves as a benchmark scenarios that can be probed by different neutrino scattering experiments using different neutrino sources~\cite{CONNIE:2019xid,CONUS:2021dwh,Karadag:2025muq}.
In the universal model, the charges are equal to $Q'_\ell \equiv Q'_f = 1$, and the coupling becomes the same for all the fermions.
In Fig.~\ref{fig:LMlimit}, we show the constraints obtained in this sensitivity study compared to the constraint from the combined analysis of current \cenns experiments reported in Ref.~\cite{AtzoriCorona:2025ygn}, from which one can notice that future detectors will have the capability of strengthening the current constraints especially for $M_{Z'}\gtrsim0.1\, \mathrm{MeV}$. Moreover, we find that considering the two different systematic uncertainties on the CE$\nu$NS signal in the COH-Ar-750 analysis does not lead to a noticeable change in the resulting constraints.
It is interesting to notice that the contribution from the neutrino-electron scattering process, which can only be considered for CsI detectors, it only marginally improves the constraints, as shown by the small second bump at light $Z'$ masses.

\section{Conclusions}
\label{sec:conclusions}

In this work, we present a comprehensive sensitivity study of future CE$\nu$NS detectors under development by the COHERENT Collaboration.
We focus on two different experimental setups: a cryogenic cesium-iodine detector (COH-CryoCsI) and a tonne-scale liquid argon one (COH-Ar-750). We show that these detectors are expected to significantly enhance the precision of key electroweak and neutrino parameters, both within the SM and in extensions beyond it.
In particular, our study proves that next-generation COHERENT detectors will enable precise measurements of the weak mixing angle at low energies, helping to fill the gap between high-energy and low-energy determinations. Systematic uncertainties remain the dominant limitation for precision tests. Therefore, the development of future detectors must be accompanied by the effort to determine precisely the neutrino flux with the $\rm D_2O$ detector, which will be essential for extracting such fundamental parameters. Additionally, these detectors will allow for accurate measurements of the neutron distribution radius in the nuclei of cesium, iodine and argon, with COH-CryoCsI II expected to reach a precision of about 1\%.
We also demonstrate that with future CE$\nu$NS detectors it will be possible to resolve existing degeneracies in the neutrino charge radius determinations, providing constraints comparable to or better than those from current global fits. Moreover, we show that it will be possible to set competitive bounds, although not yet leading, on neutrino electromagnetic properties beyond the standard model, such as magnetic moments and millicharge.
Finally, we investigate the potential to detect new light mediators that would indicate nonstandard neutrino interactions. The proposed detectors will strengthen existing constraints on the mass and coupling of such hypothetical particles.
Since its first detection, \cenns has now entered a new phase of precision. The upcoming cryogenic CsI and tonne-scale liquid argon detectors will be fundamental in advancing our understanding of electroweak interactions at low energies.

\begin{acknowledgements}
We would like to thank D. Pershey for providing useful information and details regarding the sensitivity of the COHERENT CryoCsI detector.
\end{acknowledgements}

\bibliography{ref}

\end{document}